\title{Uncoupled Learning Rules for Seeking Equilibria\\ in Repeated Plays: An Overview}
\author{M. Sadegh Talebi\\
\small Department of Automatic Control\\
\small School of Electrical Engineering\\
\small KTH The Royal Institute of Technology\\ 
\small \textit{email: mstms@kth.se}
}

\date{}

\documentclass[11pt,a4paper]{article}

\usepackage{amssymb}
\usepackage{amsmath} 
\usepackage{graphicx}   % need for figures
\usepackage{verbatim}   % useful for program listings
\usepackage{color}      % use if color is used in text
\usepackage{subfigure}  % use for side-by-side figures
\usepackage{stmaryrd}
\usepackage{bbm}
\usepackage{tabularx}
\usepackage{pifont}
\usepackage{multirow}

\newtheorem{myDef}{Definition}

\newtheorem{myTheo}{Theorem}

\begin{document}
\maketitle

\begin{abstract}
In this note, we consider repeated play of a finite game using learning rules whose period-by-period behavior probabilities or empirical distributions converge to some notion of equilibria of the stage game. Our primary focus is on uncoupled and completely uncoupled learning rules. While the former relies on players being aware of only their own payoff functions and able to monitor the action taken by the others, the latter assumes that players only know their own past realized payoffs. 
We highlight the border between possible and impossible results using these rules. We also overview several uncoupled and completely uncoupled learning rules, most of which leverage notions of regret as the solution concept to seek payoff-improving action profiles.  
\end{abstract}

%%%%%%%%%%%%%%%%%%%%%%%%%%%%%%%%%%%%%%%%%%%%%%%%%%%%%%%%%%%%%%%%%%%%%%%%%%%%%%%%%%%%%%%%%%%%%%%%%%%%%%%%%
%%%%%%%%%%%%%%%%%%%%%%%%%%%%%%%%%%%%%%%%%%%%%%%%%%%%%%%%%%%%%%%%%%%%%%%%%%%%%%%%%%%%%%%%%%%%%%%%%%%%%%%%%
%%%%%%%%%%%%%%%%%%%%%%%%%%%%%%%%%%%%%%%%%%%%%%%%%%%%%%%%%%%%%%%%%%%%%%%%%%%%%%%%%%%%%%%%%%%%%%%%%%%%%%%%%

\section{Introduction}
One of active and interesting research areas in game theory addresses rules and procedures to find equilibria in a repeated play of a game. This area has received many attentions in the course of the last two decades both in economics (e.g., \cite{fudenberg_levine,HM_2013} and references therein) and different disciplines of engineering (e.g., \cite{nash_seeking, nash_seeking_2, krishna}). 

In this paper, we consider repeated play of a finite game and overview learning rules whose period by period behavior probabilities or empirical distributions converge to some notion of equilibria of the underlying stage game. So far a plethora of such rules has been investigated and they can be categorized with respect to their degree of rationality and cognitive optimization.  

A large class of such learning rules constitutes \emph{simple} and \emph{natural} rules, also referred to as ``adaptive heuristics'' (see e.g. \cite{HM_2013}), whose rationality requirement is less than that of complicated rules (i.e. those that require decision making via certain Bayesian beliefs) and also higher than that of evolutionary dynamics. 
By \emph{simplicity} we require the rule to be implemented easily (in terms of time and computation complexity, etc.), which implies that it can be implemented by players who have \emph{bounded rationality}. 
By \emph{natural} we mean that the learning rule should be devised for a natural situation in which each  player can monitor the actions taken by other players and knows her payoff function but not the payoff of the others. This setup, to be defined precisely later, is called \emph{uncoupledness}. Yet, there is another setup, \emph{complete uncoupledness}, which is more confined and preserves the privacy of players in the sense that no player can observe the action taken by the others. This setup is sometimes called ``the case of unknown game'' as each player views the game as a black box. This setting is closely related to adversarial (nonstochastic) multi-armed bandit problem (see e.g. \cite{nonstochastic_bandit, LPG}) in which a player, who plays against an adversary, is unaware of the payoff functions involved in the game (including her own payoff) as well as the action taken by the adversary. What she knows is her choice of actions and her own earned payoffs.    

Although sophisticated rules might yield a more elegant behavior (in terms of convergence speed, efficiency of equilibria, etc.) in engineering disciplines as they might be exploited by smart players, only hardly could such rules capture the underlying models in game-theoretic interactions between humans, mutants, etc. It should be emphasized, however, that even in engineering disciplines, there is the curse of dimensionality with sophisticated rules and indeed many of them cannot be realized even by smart players. Amongst such rules one can mention those that require the entire history of play to be taken into account when playing the game. Such a rule may be realized using some automata whose number of states can grow without bound as the game progresses.

%%%%%%%%%%%%%%%%%%%%%%%%%

In this overview paper, we address simple and uncoupled learning rules that when employed for a repeated play of game, their period-by-period behavior probabilities or empirical distributions converge (in some sense) to some notion of equilibria of the stage game. Towards this, an important question that comes to mind is: Is it possible to find Nash equilibria using uncoupled (and completely uncoupled) rules in every multi-player game? Unfortunately the answer is negative \cite{HM_impossibility}, and therefore a fundamental step when studying such rules is to distinguish between possible and impossible results. We then have to seek a less ambitious goal (e.g. seeking correlated equilibria) or to relax the underlying assumption (e.g. to allow limited coordination or confine the result to some games but not all). The good news is that there are uncoupled learning rules that can reach correlated equilibria in every game. Having all these said, one may deduce that some coordination should be involved in either the solution concept or the learning rule.

Most of the learning rules to be considered here employ some notion of regret, perhaps to be treated as their underlying solution concept, which tries to capture the degree of cumulative dis-satisfaction of a player had she chosen another action. The very idea behind such regret-based rules is to try alternative actions in some payoff-improving directions, to be signaled by the regret. The usefulness of regret is that it yields a simple and natural way of playing as it relies on a player who is (completely) unaware of the other players' payoff function. Regret minimizing rules constitute a large category of rules and we only address here some of them that can reach equilibria. The concrete theory behind regret minimization is mainly built on Hannan-consistency and Blackwell's approachability, which seem to be beyond the scope of this paper and thereby will be pursued here only partially. 

Before giving the organization of the paper, we emphasize that we do not consider special types of games such as potential games, supermodular games, etc. Rather, we consider general finite multi-player games and sometimes restrict our focus to generic multi-player games.  
As we already mentioned, the main concentration here is on learning rules that converge\footnote{Convergence types and notions will be defined in detail in the next section.} to (some notions of) equilibria. We stress that this is different from the problem of determining whether the steady state point of some learning rule in the game of interest is an equilibrium.

\paragraph{Organization of the paper.} In Section \ref{sec:model}, we describe the game model and some definitions regarding the notions of equilibria and convergence. In Section \ref{sec:regret}, we give some definitions as well as results for regret and state some results for achieving correlated equilibria in general games. Section \ref{sec:RM} and Section \ref{sec:RT} describe some regret-based learning rules for correlated equilibria and mixed equilibria, respectively. Section \ref{sec:TEL} is devoted to explain important learning rule for pure equilibria. 
Section \ref{sec:fic_play} briefly describes \emph{fictitious play}. In Section \ref{sec:poss_imposs}, we distinguish between possibility and impossibility results for both uncoupled and completely uncoupled learning rules. Finally, we discuss some benefits and drawbacks of described learning rules as well as some open problems in Section \ref{sec:discuss}.

%%%%%%%%%%%%%%%%%%%%%%%%%%%%%%%%%%%%%%%%%%%%%%%%%%%%%%%%%%%%%%%%%%%%%%%%%%%%%%%%%%%%%%%%%%%%%%%%%%%%%%%%%%%%
%%%%%%%%%%%%%%%%%%%%%%%%%%%%%%%%%%%%%%%%%%%%%%%%%%%%%%%%%%%%%%%%%%%%%%%%%%%%%%%%%%%%%%%%%%%%%%%%%%%%%%%%%%%%
%%%%%%%%%%%%%%%%%%%%%%%%%%%%%%%%%%%%%%%%%%%%%%%%%%%%%%%%%%%%%%%%%%%%%%%%%%%%%%%%%%%%%%%%%%%%%%%%%%%%%%%%%%%%

\section{Model of Game}
\label{sec:model}
We consider the finite normal form game represented by the triplet $G=\langle N,S,\pi\rangle$ with the mixed action\footnote{Here, we use the term \emph{strategy} for the repeated game and the term \emph{action} for the stage game.} extension $\tilde G=\langle N,\boxdot(S),\tilde \pi\rangle$. 
The set $N=\{1,2,\dots,n\}$ denotes the set of players. We assume that each player $i$ has a finite set of pure actions $S_i=\{1,2,\dots,m_i\}$. The $n$-vector $s=(s_1,\dots,s_n)$ where $s_i\in S_i$, is a pure action profile and the Cartesian product $S=\varotimes_{i\in N} S_i$ denotes the pure action space of the game. 

Let $\pi_i:S\rightarrow \mathbb R$ be the pure-action payoff function (or simply the payoff) of player $i$. Here, we assume that payoffs are bounded, i.e. there exists some constant $M$ such that $|\pi_i(s)|\leq M$ for all $i\in N$ and $s\in S$. 
Then, $\pi:S\rightarrow \mathbb R^n$ denotes the combined pure-action payoff function. 

The set of mixed actions of player $i$ is $\Delta(S_i)$ which is a unit simplex in $\mathbb R^{m_i}$ where a mixed action $x_i=(x_{ih},h\in S_i)\in \Delta(S_i)$ is a probability vector. The vector $x=(x_1,x_2,\dots,x_n)$ with $x_i\in \Delta(S_i)$, is the mixed action profile and the set $\boxdot(S)=\varotimes_{i\in N} \Delta(S_i)$ is the mixed action space of the game.    

Besides Nash equilibria and correlated equilibria, we are also interested in Nash $\epsilon$-equilibria which is defined next.

\begin{myDef} \cite{LPG,HM_2006}
The pure action $s_i\in S_i$ is said to be an $\epsilon$-best reply (with $\epsilon>0$) to $s_{-i}$ if 
$$\pi_i(s_i,s_{-i})\geq \pi_i(s^\prime_i,s_{-i})-\epsilon,\quad \forall s^\prime_i\in S_i.$$ 
Similarly, $x_i\in\Delta(S_i)$ is an $\epsilon$-best reply to $x_{-i}$ if $$\tilde\pi (x_i,x_{-i})\geq \tilde \pi(y_i,x_{-i})-\epsilon, \quad \forall y_i\in \Delta(S_i).$$  
The mixed action profile $x$ is a Nash $\epsilon$-equilibrium if for every $i\in N$, $x_i$ is an $\epsilon$-best reply to $x_{-i}$.
\end{myDef}

It is worthwhile to highlight that in contrast to pure and mixed Nash equilibria that are exact points, a mixed Nash $\epsilon$-equilibrium is not an exact point; indeed it constitutes a set \cite{babi_2012}. 

\subsection{Repeated Game Setup}
We denote by $x_i^t\in \Delta(S_i)$ the mixed action of player $i$ at time $t$ and by $s_i^t\in S_i$ the \emph{realized} pure action at time $t$. Similarly $x^t=(x_1^t,\dots,x_n^t)\in \boxdot(S)$ and $s^t=(s_1^t,\dots,s_n^t) \in S$ represent the mixed action profile at time $t$ and the realized pure action profile at time $t$, respectively. 
The history of play up to time $t$ is denoted by $H_t=(s^1, s^2, \dots, s^t)$, where $s^{\tau}\in S$ is the realized action profile at time $\tau\leq t$.   

We let $o_i^t$ be the \emph{observation sequence} of player $i$ up to time $t$ that denotes the sequence of actions and payoffs she \emph{knows or can observe} up to time $t$. 
For instance, for the case of uncoupled learning rules discussed in Section 1, the observation sequence of each player $i$ takes the form $o_i^t=(\{s^\tau\}_{\tau=1,\dots,t},\pi_i(.))$. 
Also let $O_i^t$ denote the set of all possible observation sequences of player $i$ up to time $t$. 

The strategy of player $i\in N$ is a sequence of functions $f_i=(f^1_i,f^2_i,\dots,f^t_i,\dots)$ where for each $\tau$, $f_i^\tau$ is a mapping $f_i^\tau:O_i^{\tau-1}\rightarrow \Delta(S_i)$ that assigns a mixed action to every possible observation sequence of player $i$. Moreover, $f=(f_1,\dots,f_n)$ denotes the strategy profile.

\subsection{Models of Learning Rules}
Here we give some definitions for learning rules we consider in subsequent sections. 

\begin{myDef}\cite{HM_2006}
A learning rule is said to be $R$-recall if every player conditions her play on the last $R$ periods of play. A strategy is said to be \emph{finite recall} if such a positive integer $R$ exists. Then for each $t>R$, the function $f_i^t$ admits the form $f^t_i(o_i^{t-1}\backslash o_i^{t-R-1})$. 
It is moreover \emph{stationary} if time $t$ doesn't matter, i.e. $f_i^t\equiv f_i(o_i^{t-1}\backslash o_i^{t-R-1})$. 
\end{myDef}

\begin{myDef}
A learning rule is said to be \emph{finite memory} if each player can implement it by a finite automata. An $R$-memory learning rule is a finite memory rule such that its automata has $|S|^R$ states \cite{babi_2010, HM_2006}.
\end{myDef}

Similar to finite-recall rules, finite memory rules condition play on finitely many periods of history of play and therefore finite-recall rules constitute a subclass of finite memory rules. 
However, finite memory rules do not necessarily use the last periods of history, and hence their recall can grow without bound \cite{babi_2010}.   

Implication of possibility of realization of finite memory rules by finite automata is that they can be modeled by a Markov chain with finitely many states. It's worth mentioning that finite-memory is a significant desired property. By contrast, infinite-memory rules are problematic as a player who wishes them has to exploit more and more data as play proceeds. Therefore, having finite-memory property is a natural requirement.

\subsubsection{Uncoupled Learning Rules}

\begin{myDef}
A learning rule is said to be \emph{uncoupled} if the strategy of each player $i$ 
depends on the payoff function of herself and the history of play of all other players.
\end{myDef}

The definition above implies that for uncoupled learning rules, the observation sequence of each player $i$ takes the form $o_i^t=(\{s^\tau\}_{\tau=1,\dots,t},\pi_i(.))$. 
Consequently, the mixed action of player $i$ for time $t$ is given by $x_i^t=f_i^t(o_i^{t-1})=f_i^t(\{s^\tau\}_{\tau=1,\dots,t-1},\pi_i)$. 

The class of uncoupled learning rules suits decentralized systems where each node is unaware of the preference of other players\footnote{Note that this setup is more restrictive than the one in which a player knows the payoff structures involved but not the earned (realized) payoffs of the others. Indeed, in uncoupled setup, each player knows nothing about the payoff structure of the others.}. 
This class of rules is rather wide. Perhaps the most well-known uncoupled learning rule is 
replicator dynamics. However, many other interesting rules are uncoupled amongst which are: best-reply, fictitious play, etc. 

\subsubsection{Completely Uncoupled Learning Rules}
One important subclass of uncoupled rules is the class of completely uncoupled rules defined next.

\begin{myDef}
A learning rule is said to be \emph{completely uncoupled\footnote{Also referred to as \emph{radically uncoupled} in \cite{regret_testing}.}} or \emph{payoff-based} if the strategy of each player only depends on her own realized payoffs and her own past actions. 
\end{myDef}

In this setting, the observation sequence of each player $i$ is $o_i^t=(\{s_i^\tau,\pi_i(s^\tau)\}_{\tau=1,\dots,t})$. 
Consequently, the mixed action of player $i$ for time $t$ is given by $x_i^t=f_i^t(o_i^{t-1})=f_i^t(\{s_i^\tau,\pi_i(s^\tau)\}_{\tau=1,\dots,t-1})$. 
In the case of completely uncoupled rules, each player doesn't know so much about the game. Indeed, she doesn't know the number of players involved in the game, let alone the history of their plays. Even she doesn't know her own payoff function completely. She only has knowledge about her \emph{realized payoffs}, i.e. payoff values she has received so far. In this setting, she views the game as a black box where she chooses an action and receives a payoff, and that's why these rules are sometimes called \emph{the case of unknown game}. 

Some completely uncoupled learning rules have been proposed so far; two well-known such rules are \texttt{Trial-and-Error Learning}\footnote{To avoid confusion, particular learning rules are typeset using \texttt{courier font}.} \cite{LTE} and (a variant of) \texttt{Regret Testing} \cite{regret_testing}.

\subsection{Notions of Convergence to Equilibria}
Here we introduce notions of convergence that will become useful when addressing learning rules in later sections. Some of the following notions deal with the convergence of empirical distributions of play to some equilibrium (or a set of equilibria) whereas the others imply that the behavior of the game (or average behavior over bounded time intervals) comes close to some equilibrium (or a set of equilibria). The formal definition of these notions have been adopted from \cite{Strategic_learning}.  

Before proceeding to define these notions, we mention three types of convergence. 
Let $Y_1,Y_2,\dots,Y_t$ be random variables in the same probability space $(\Omega,\mathcal F, P)$. Then, 
\begin{itemize}
\item $Y_t$ is said to converge to $Y$ in probability if for all $\epsilon>0$, 
$$\lim_{t\rightarrow\infty}P\left(|Y_t-Y|\geq \epsilon\right)=0.$$
\item $Y_t$ is said to converge to $Y$ \emph{almost surely} if 
$$P\left(\lim_{t\rightarrow\infty} Y_t=Y\right)=1.$$
\item $Y_t$ is said to converge to the set $\mathcal Y$ with frequency $1-\epsilon$ if $\{Y_\tau\}_{\tau=1}^\infty$ belongs to the set $\mathcal Y$ with frequency $1-\epsilon$ or equivalently if $$\lim_{t\rightarrow\infty}\inf\frac{|\{1\leq \tau\leq t: Y_{\tau} \in \mathcal Y\}|}{t}\geq 1-\epsilon,\quad \hbox{almost surely}.$$
\end{itemize}
Note that among the others, almost sure convergence is the strongest notion.

Now we proceed to define the notions of convergence for learning rules. 
For each $s\in S$, let $\Phi_t[s]=\frac{1}{t}|\{1\leq \tau\leq t:s^\tau=s\}|$. Then \emph{the empirical joint distribution of play}, also known as \emph{long run sample distribution of play} and \emph{long run empirical behavior}, is defined by 
$$(\Phi_t[s])_{s\in S}\in\boxdot(S)\footnote{Similarly, the empirical marginal distribution of play of player $i$ is defined as follows 
$$(\Phi_t[s_i])_{s_i\in S_i}\in\Delta(S_i),$$ where $\Phi_t[s_i]=\frac{1}{t}|\{1\leq \tau \leq t:s_i^\tau=s_i\}|$.}.$$

Let's consider $\Phi_t$ as a vector in $\boxdot(S)$ whose $s$th component is $\Phi_t[s]$. Let $\Delta^*$ be a non-empty and closed subset of $\boxdot(S)$. Then, 
\begin{itemize}
\item Long run empirical behavior (or for short, empirical distribution) of play is said to converge to $\Delta^*$ if \mbox{$\textrm{\textbf{dist}}(\Phi_t,\Delta^*)\rightarrow 0$} almost surely\footnote{The function $\textrm{\textbf{dist}}(z,A)$ denotes the Euclidean distance between point $z$ and set $A$.}.  
\item Long run empirical behavior (empirical distribution) of play is said to converge pointwise to $\Delta^*$ if there exists some $q\in \Delta^*$ such that $\Phi_t\rightarrow q$ almost surely.  
\end{itemize}

These notions of convergence are rather weak and do not imply that behavior or actual play at any given point or even over a finite time interval is close to the desired equilibrium set. 
A more demanding and also desirable notion is that in the short run, behavior probabilities come close to equilibrium. In this case indeed we are interested in the convergence of players' behavioral strategies, instead of convergence of realizations of those strategies.  
Now consider the sequence of conditional probability distributions $(q^1,q^2,\dots,q^t,\dots)$ where  $q^t=q^t(.|H_t)\in \boxdot(S)$ is the distribution of joint actions at time $t$ conditional on the history $H_t$. Then, 
\begin{itemize}
\item Period-by-period behavior probabilities of play is said to converge to $\Delta^*$ if \mbox{$\textrm{\textbf{dist}}(q^t,\Delta^*)\rightarrow 0$} almost surely.  
\item Period-by-period behavior probabilities of play is said to converge pointwise to $\Delta^*$ if there exists some $q^*\in \Delta^*$ such that $q^t\rightarrow q^*$ almost surely.  
\end{itemize}

In the sequel, ``behavior probabilities'' will be used to denote ``period-by-period behavior probabilities''. It is worth mentioning that in behavior-related notions, what is important is the convergence in terms of players' strategic intentions, i.e. conditional probability distributions, not the realization of those intentions. Finally, it should be emphasized that convergence of behavior probabilities of play is only relevant for the case of mixed and correlated equilibria.

\section{Mixed and Correlated Equilibria: Solution Concepts}
\label{sec:regret}
As mentioned before, a plethora of uncoupled learning rules aim at minimizing some notion of the regret or at least takes the advantage of regret as an indicator towards finding payoff-improving action(s). The key idea behind this solution concept is to bring the cumulative loss in payoff to the minimum.  

This section is devoted to the definition of some notions of regret and their connection to equilibria in simple setups. The simplest notion of regret, referred to as the cumulative regret or simply regret, reflects the cumulative dissatisfaction of a player had she chosen constantly some specific action. Formally for each player $i$, we define the regret with respect to action $j\in S_i$ up to time $t$ as  
$$r^i_{t,j}=\sum_{\tau=1}^t \pi_i(j,s_{-i}^\tau)-\sum_{\tau=1}^t \pi_i(s^\tau).$$
Having defined this, now we can define the internal regret of a player. 

The definition of regret above suggests its uncoupled nature, as only the own payoff function of the player  is needed for regret computation. Although regret admits a simple definition, as we will see in Section \ref{sec:RT}, it can be used to find approximate mixed equilibria of almost all games.

\subsection{Internal Regret}
\begin{myDef}
For each player $i$ and pair of actions $j,j^\prime$, the internal regret up to time $t$ is defined as her payoff loss had she chosen action $j^\prime$ every time she played action $j$, i.e.   
$$R^{\textrm{int},i}_t(j,j')=\sum_{\tau\leq t: s_i^\tau=j}(\pi_i(j',s_{-i}^\tau)-\pi_i(s^\tau)).$$
\end{myDef}

The concept of internal regret is powerful as it is able to capture the player's dissatisfaction more precisely compared to regret defined earlier. As a result, one may expect that internal-regret-minimizing rules, i.e. rules that achieve the minimum internal regret, to be powerful for obtaining equilibria. Internal-regret-minimizing rules have been investigated deeply for calibration and forecasting in the course of the last decade. When employed to play games, they are powerful enough to reach approximate correlated equilibria in every finite game. Indeed, we have the following result.

\begin{myTheo}\cite{LPG}
In a multi-player game, if each player plays according to an internal-regret-minimizing rule, then the distance between empirical distributions of play and the set of correlated equilibria of 
game converges to zero almost surely.
\end{myTheo}

One uncoupled internal-regret-minimizing rule  is \texttt{Regret Matching} rule which will be described in the next section. 

\section{Correlated Equilibria: Learning Rules}
\label{sec:RM}

\subsection{Regret Matching}
\texttt{Regret Matching} is an uncoupled rule which is based on the minimization of internal regret and has been proposed by Hart and Mas-Collel \cite{HM_simple_adaptive, HM_reinforcement_learning}. This rule is quite simple: at each period of play, each player establishes the vector of average internal regrets (with respect to her current action). At the next period, she then switches to another action with a probability that is proportional to the average internal regret of that action. This rule is described formally below.\\ 

\footnotesize
\begin{center}
\begin{tabularx}{.82\textwidth}{X}\
\ \\
\normalsize
\textbf{Algorithm 2:} \textbf{\texttt{Regret Matching}} \cite{HM_simple_adaptive, HM_reinforcement_learning}\\
\hline

Strategy for player $i$:\\
\ \\
\textbf{Parameters:} $\mu^i\in(2M_i(m_i-1),\infty)$ where $M_i$ is an upper bound for $|\pi_i(.)|$.\\
\ \\
\textbf{Initialization:} \\
Choose a mixed action $\pi_0^{(i)}\in \Delta(S_i)$ uniformly at random. Set $t=1$.\\
\ \\
\textbf{Loop:}\\
1. Play according to $x_i^t\in \Delta(S_i)$. Let $j=s_i^t$.\\
\ \\
2. For each pair $k,j$ compute the average of internal regret  
\begin{eqnarray}
\overline{R}^{\textrm{int},i}_t(j,k)=\frac{1}{t}\sum_{\tau\leq t: s_i^\tau=j}(\pi_i(k,s_{-i}^\tau)-\pi_i(s^\tau)).\nonumber
\end{eqnarray}
3. Compute the mixed action for the next period of play
\begin{eqnarray}
x_{ik}^{t+1}&=&\frac{1}{\mu_i}[\overline{R}^{\textrm{int},i}_t(j,k)]^+, \quad \forall k\neq j,\nonumber\\
x_{ij}^{t+1}&=&1-\sum_{k\in S_i, k\neq j}x_{ik}^{t+1}.\nonumber
\end{eqnarray}
4. Set $t=t+1$ and repeat the loop.\\

\hline
%\ \\
\end{tabularx}
\end{center}
\normalsize

For the \texttt{Regret Matching} rule we have the following result. 

\begin{myTheo}\cite{HM_simple_adaptive}
If every player plays according to \emph{\texttt{Regret Matching}} procedure, then the empirical distributions of play converge almost surely as $t\rightarrow\infty$ to the set of correlated equilibria of the stage game.  
\end{myTheo}

\begin{figure}[th]
\centering
\includegraphics[scale=1, angle=-1]{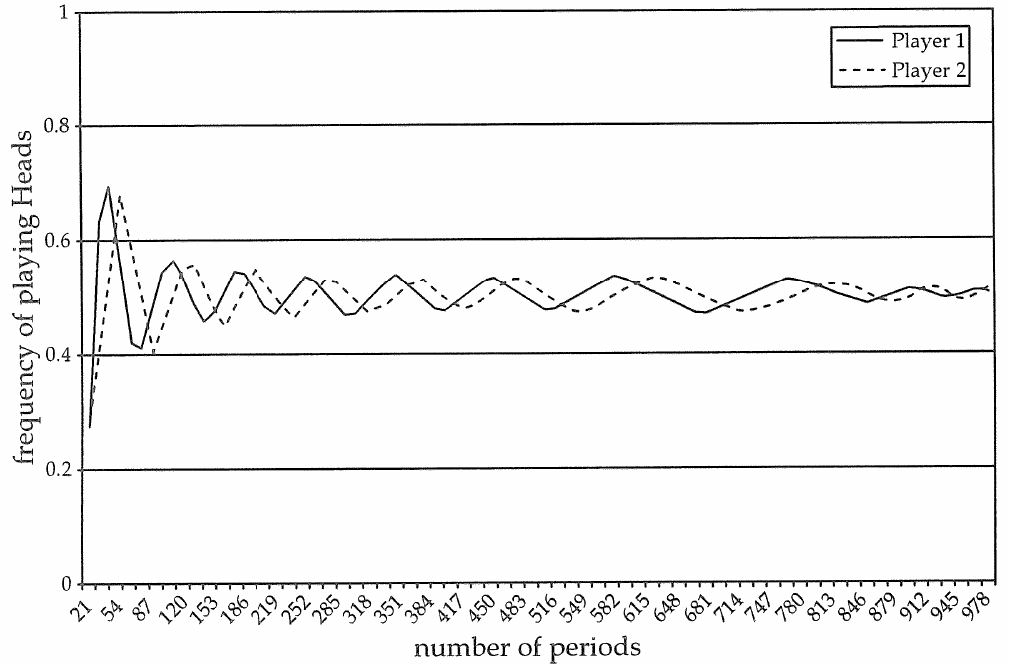}
\caption{Cumulative frequencies of \texttt{Regret Matching} in Matching Pennies game (courtesy of \cite{Strategic_learning})}
\label{fig:reg_matching_LT}
\end{figure}

\begin{figure}[th]
\centering
\includegraphics[scale=1, angle=-1]{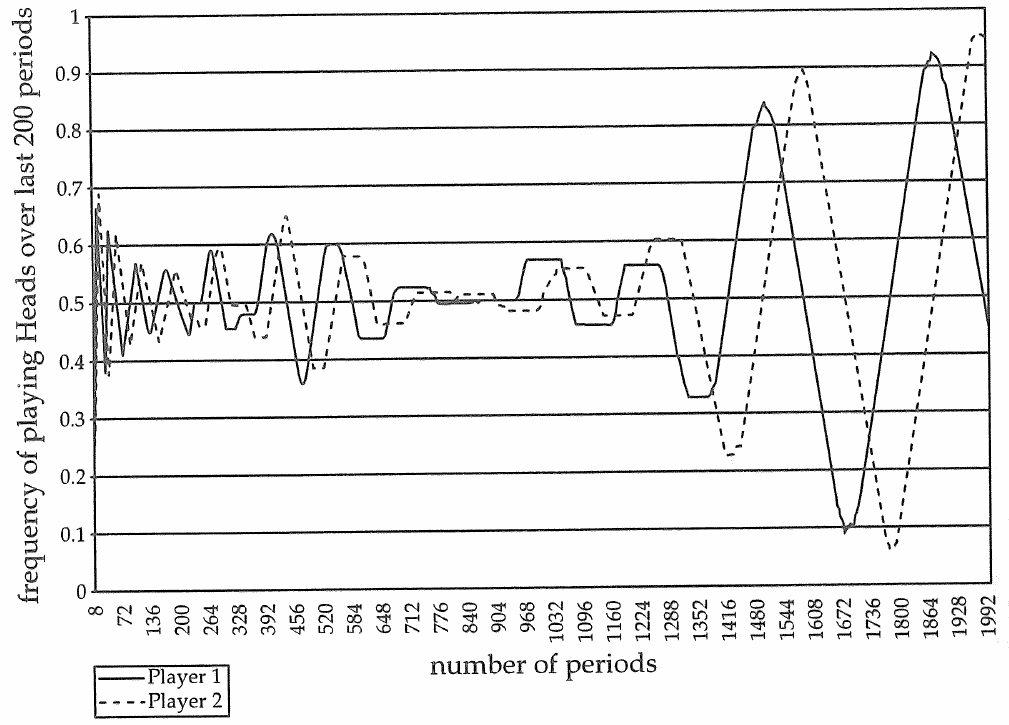}
\caption{Short-run instabilities of \texttt{Regret Matching} in Matching Pennies game where empirical averages are obtained over 100 periods (courtesy of \cite{Strategic_learning})}
\label{fig:reg_matching_ST}
\end{figure}

Figure \ref{fig:reg_matching_LT} shows two sample trajectories of joint empirical distributions of \texttt{Regret Matching} rule when applied to Matching Pennies game. This is consistent with statement of Theorem 2 and confirms that long run behavior (empirical frequencies) of this rule converges to correlated equilibria. On the other hand, behavior probabilities of \texttt{Regret Matching} is not guaranteed to converge; Figure \ref{fig:reg_matching_ST} depicts the moving average of empirical distributions for \texttt{Regret Matching} in Matching Pennies over a window of 200 periods, which is fixed window but rather long. This reveals that the short-run behavior of this rule can be quite unstable and tends to spend long intervals away from the correlated equilibrium. Therefore, \texttt{Regret Matching} might not converge in behavior probabilities.

\subsubsection{Modified Regret Matching}
\texttt{Regret Matching} rule is not completely uncoupled as it requires each player to know her own payoff function and also to track the action taken by the others. This drawback can be obviated by estimating the average internal regret as follows

$$\hat{\overline R}^{\textrm{int},i}_t(j,k)=\frac{1}{t}\sum_{\tau\leq t: s_i^\tau=k}\frac{x^\tau_{ij}}{x^\tau_{ik}}\pi_i(s^\tau)-\frac{1}{t}\sum_{\tau\leq t: s_i^\tau=j}\pi_i(s^\tau).$$
Observe that now $\hat{\overline R}^{\textrm{int},i}_t(j,k)$ is only dependent on $\pi_i(s^1),\dots,\pi_i(s^t)$.

\footnotesize
\begin{center}
\begin{tabularx}{.82\textwidth}{X}\
\ \\
\normalsize
\textbf{Algorithm 2:} \textbf{\texttt{Modified Regret Matching}}\footnote{Parameters have been adopted from the correction to \cite{HM_reinforcement_learning} published by the same authors.} \cite{HM_reinforcement_learning}\\
\hline
\footnotesize
Strategy for player $i$:\\
\ \\
\textbf{Parameters:} $\gamma\in(0,1/4)$ and $\mu^i\in(2M_im_i,\infty)$ where $M_i$ is an upper bound for $|\pi_i(.)|$.\\
\ \\
\textbf{Initialization:} \\
Choose a mixed action $\pi_0^{(i)}\in \Delta(S_i)$ uniformly at random. Set $t=1$.\\
\ \\
\textbf{Loop:}\\
1. Play according to $x_i^t\in \Delta(S_i)$. Let $j=s_i^t$.\\
\ \\
2. For each pair $k,j$ compute the vector
\begin{eqnarray}
\overline{\hat R}^{\textrm{int},i}_t(j,k)=\frac{1}{t}\sum_{\tau\leq t: s_i^\tau=k}\frac{x^\tau_{ij}}{x^\tau_{ik}}\pi_i(s^\tau)-\frac{1}{t}\sum_{\tau\leq t: s_i^\tau=j}\pi_i(s^\tau)\nonumber
\end{eqnarray}
3. Compute the mixed action for the next period of play
\begin{eqnarray}
x_{ik}^{t+1}&=&(1-\frac{\delta}{t^\gamma})\min\left(\frac{1}{\mu_i}[\hat{\overline R}^{\textrm{int},i}_t(j,k)]^+,\frac{1}{m_i}\right)+\frac{\delta}{m_it^\gamma}, \quad \forall k\neq j,\nonumber\\
x_{ij}^{t+1}&=&1-\sum_{k\in S_i, k\neq j}x_{ik}^{t+1}\nonumber
\end{eqnarray}
4. Set $t=t+1$ and repeat the loop.\\

\hline
%\ \\
\end{tabularx}
\end{center}
\normalsize

For the \texttt{Modified Regret Matching} rule we have the following result. 

\begin{myTheo}\cite{HM_reinforcement_learning}
If every player plays according to the  \texttt{\emph{Modified Regret Matching}} procedure, then the empirical distributions of play converge almost surely as $t\rightarrow\infty$ to the set of correlated equilibria of the stage game.  
\end{myTheo}

\section{Mixed Equilibria: Learning Rules}
\label{sec:RT}
In this section, we describe three variants of a procedure called \texttt{Regret Testing}, which leverages regret as its underlying solution concept. 
\texttt{Regret Testing} was originally proposed for two-player games by Foster and Young in \cite{regret_testing}. In their paper, Foster and Young showed that this procedure leads to play of Nash equilibria with frequency 1 in every 2-player game and also presented uncoupled and completely uncoupled variants of it. 

Following that work, Germano and Lugosi in \cite{regret_testing_germano} proposed three learning rules as extensions of \texttt{Regret Testing} procedure which work in any generic multi-player game. These rules are \texttt{Experimental Regret Testing} and \texttt{Annealed Localized Experimental Regret Testing} abbreviated here as \texttt{ALERT}, and \texttt{Payoff-based ALERT}. Moreover, they proved that the three variants of \texttt{Regret Testing} can indeed yield almost sure convergence of play to Nash equilibria which is stronger than convergence with frequency 1. As we will see in the sequel, \texttt{Experimental Regret Testing} is not uncoupled as it requires (uncoupled) additional information of the game to be known by the players. By contrast, \texttt{ALERT} is an uncoupled rule and \texttt{Payoff-based ALERT} is completely uncoupled.  
In subsequent subsections, we describe the three aforementioned variants of  \texttt{Regret Testing} along with some convergence results.
 
\subsection{Experimental Regret Testing}
The  \texttt{Experimental Regret Testing}  procedure works as follows. Time is partitioned into frames, each comprising  $T$ periods. At the beginning of each frame, each player $i$ chooses a mixed action from $\Delta(S_i)$. Then she plays according to this mixed action during the frame (i.e. $T$ periods). Then, at the end she computes the vector of average regrets. 
If one of the elements of this vector exceeds $\rho$, she draws a new action uniformly at random from $\Delta(S_i)$. On the other hand, if none of the elements of this vector exceeds $\rho$, then, with probability $1-\lambda$, she continues playing according to the previous action for the next $T$ periods (the next frame), and with probability $\lambda$, she draws a new mixed action from $\Delta(S_i)$. 
A more formal description of \texttt{Experimental Regret Testing} is depicted below. Note that choosing $\lambda=0$, the following procedure degenerates to the \texttt{Regret Testing} procedure proposed by Foster and Young in \cite{regret_testing}. \\

\footnotesize
\begin{center}
\begin{tabularx}{.82\textwidth}{X}\
\ \\
\normalsize
\textbf{Algorithm 3:} \textbf{\texttt{Experimental Regret Testing}} \cite{regret_testing_germano}\\
\hline
Strategy for player $i$:\\
\ \\
\textbf{Parameters:} Triple $(T,\rho,\lambda)$, where $T\in\mathbb N$, $\rho>0$, and $\lambda\in(0,1).$\\
\ \\
\textbf{Initialization:} \\
Choose a mixed action $\pi_0^{(i)}\in \Delta(S_i)$ uniformly at random. Set $t=1$.\\
\ \\
\textbf{Loop:}\\
1. Play according to $x_i^t\in \Delta(S_i)$ for $T$ periods.\\
\ \\
2. Compute the vector of average regrets over $T$ periods, i.e. the vector with elements 
\begin{eqnarray}
\overline{r}^{i}_{t,k}=\frac{1}{T}\sum_{\tau=t+1}^{t+T}\left(\pi_{i}(k,s_{-i}^\tau)-\pi_i(s^\tau)\right), \quad k\in S_i.\nonumber
\end{eqnarray}
3. If $\overline{r}^i_{t,k}\geq \rho$ for some $k\in S_i$, then for the next period select $x_i^{t+T}\in \Delta(S_i)$ uniformly at random. If $\overline{r}^i_{t,k}< \rho$ for all $k\in S_i$, then with probability $1-\lambda$, set $x_i^{t+T}=x_i^t$, and with probability $\lambda$, choose $x^{t+T}_i\in \Delta(S_i)$ uniformly at random.\\
\ \\
4. Set $t=t+T$ and repeat the loop.\\

\hline
%\ \\
\end{tabularx}
\end{center}
\normalsize

The very idea of this procedure and also the procedure proposed by Foster and Young in \cite{regret_testing} is that after not a very long search period, the mixed profile $x^{t}$ will be a Nash $\epsilon$-equilibrium. At that time, since all players would have small regret, the procedure will get stuck with this value. For \texttt{Experimental Regret Testing} the following result holds:

\begin{myTheo}\cite{regret_testing_germano}
Let $G$ be a generic $N$-player normal form game whose payoff functions have the range $[0,1]$. There exists a positive number $\epsilon_0$ such that for all $\epsilon<\epsilon_0$ there exist constants $c_1,\dots,c_4$ such that if \texttt{\emph{Experimental Regret Testing}} procedure is used with parameters 
$$\rho\in(\epsilon,\epsilon+\epsilon^{c_1}),\quad\lambda\leq c_2\epsilon^{c_3},\quad\textrm{and}\quad T\geq-\frac{1}{2(\rho-\epsilon)^2}\log(c_4\epsilon^{c_3}),$$
then for all $B\geq \frac{\log(\epsilon/2)}{\log(1-\lambda^n)}$, we have $$\Pr\left(x^{BT}\notin X_{\textrm{NE}}^\epsilon\right)\leq \epsilon,$$
where $X_{\textrm{NE}}^\epsilon$ denotes the set of Nash $\epsilon$-equilibria. 
\end{myTheo}

It is worth mentioning that this theorem guarantees the almost sure convergence of joint empirical distributions of play and does not assert anything regarding convergence of behavior probabilities. Also note that in \cite{regret_testing_germano}, another result strengthens the above to the pointwise convergence of joint empirical distributions of play to a mixed action profile that is in the convex hull of Nash $\epsilon$-equilibria. 

The analysis in \cite{regret_testing_germano} reveals some drawbacks of the \texttt{Experimental Regret Testing}. The first drawback is that the speed of convergence is quite low. Indeed, $O(\frac{1}{\epsilon^C})$ rounds are required to reach the convergence (with probability $1-\epsilon$), where $C$ is proportional to $\sum_{i\in N} m_i$, i.e. proportional to the number of pure actions in the game. Hence the convergence speed is exponentially slow as a function of $n$ and $m_i$ for each player $i$. By contrast, as will be discussed in Section \ref{sec:poss_imposs}, we can achieve correlated equilibria in an uncoupled way considerably faster.    
The second drawback is that parameters $(T,\rho,\lambda)$ depend on $\epsilon$ and on the properties of the overall game which might be unknown to all players. This is in sharp contrast to the requirement of uncoupledness where these parameters should only depend on $\epsilon$.

\subsection{Annealed Localized Experimental Regret Testing (ALERT)}

In order to resolve the second drawback of \texttt{Experimental Regret Testing}, we consider its variant called  
\texttt{Annealed Localized Experimental Regret Testing (ALERT)}. The idea behind this procedure is to anneal parameters $(T,\rho,\lambda)$. Towards this, the procedure starts with parameters $(T_1,\rho_1,\lambda_1)$ and will continue for $M_1$ frames. Then all players somehow change the parameters to $(T_2,\rho_2,\lambda_2)$ and play for $M_2$ frames, and so on. If the parameters $(T_l,\rho_l,\lambda_l)$ and $M_l, l=1,2,\dots$ will not depend on the parameters of the game, we obtain an uncoupled rule. 

Assume that $\{\epsilon_l\}_{l\in \mathbb N}$ is a decreasing sequence such that $\epsilon_l>0,l\in\mathbb N$ and $\sum_{l=1}^\infty\epsilon_l<\infty$. For brevity, we choose $\epsilon_l=2^{-l}$. Then define 
$$\lambda_l=\epsilon_l^l,\quad\rho_l=\lambda_l+\epsilon_l, \quad T_l=\left\lceil-\frac{1}{2\lambda_l^2}\log \lambda^l\right\rceil,\quad M_l=2\left\lceil\frac{\log\frac{2}{\epsilon_l}}{\log\frac{1}{1-\lambda_l}}\right\rceil.$$
Also let $x_i^{[l]}$ denote the mixed action played by player $i$ at the end of $(l-1)$st regime, and $D_\infty^i(x_i,\delta)$ be $l_\infty$-ball of radius $\delta$ centered at $x_i$. 
This procedure is described as follows.

\footnotesize
\begin{center}
\begin{tabularx}{.82\textwidth}{X}\
\ \\
\normalsize
\textbf{Algorithm 4:} \\
\textbf{\texttt{\normalsize Annealed Localized Experimental Regret Testing (ALERT)}} \cite{regret_testing_germano}\\
\hline
\footnotesize
Strategy for player $i$:\\
\ \\
\textbf{Parameters:} $\{(T_l,\rho_l,\lambda_l),M_l\}_{l\in\mathbb N}$, where $T_l,M_l\in\mathbb N$, $\rho_l>0$, and $\lambda_l\in(0,1)$, and all of them are dependent on positive and absolutely summable sequence $\{\epsilon_l\}_{l\in\mathbb N}$.\\
\ \\
\textbf{Initialization:} \\
Choose a mixed action $\pi_0^{(i)}\in \Delta(S_i)$ uniformly at random. Set $t=1$.\\
\ \\
\textbf{Loop:}\\
There are different regimes indexed by $l=1,2,\dots$. In the $l$th regime, play according to loop of \texttt{Experimental Regret Testing} with parameters $(T_l,\rho_l,\lambda_l)$ during $M_l$ periods (of length $T_l$). Instead of step 3 of \texttt{Experimental Regret Testing}, choose $x_i^{t+T_l}\in X_i$ as follows:\\
\ \\
3a. Let $\overline{r}_t^i=\max_{k\in S_i} \overline{r}_{t,k}^i$.\\
\ \\
3b. If $\overline{r}_t^i\geq \epsilon_l^{2/3}$, then select $x_i^{t+T_l}$ uniformly at random.\\
\ \\
3c. If $\rho_l\leq \overline{r}_i^t<\epsilon_l^{2/3}$, then select $x_i^{t+T_l}$ uniformly at random if, for some $t<t^\prime$ of the $l$th regime, $x_i^{t^\prime+T_l}$ has been randomly and uniformly, and otherwise select $x_i^{t+T_l}$ uniformly at random over $D_\infty^i(x_i^{[l]},\sqrt{\epsilon_l})$.\\
\ \\
4. Set $t=t+T$ and repeat the loop.\\

\hline
%\ \\
\end{tabularx}
\end{center}
\normalsize
 
Clearly, the procedure above is uncoupled since involved parameters only depend on $\{\epsilon_l\}_{l\in\mathbb N}$. For \texttt{ALERT} we have the following result. 

\begin{myTheo} \cite{regret_testing_germano}
Assume that all payoff functions in the generic $n$-player $G$ admit the compact support $[0,1]$. If each player plays according to \texttt{ALERT} with sequence $\{2^{-l}\}_{l\in\mathbb N}$, the sequence of mixed action profiles converges almost surely and 
$$\lim_{t\rightarrow\infty} x^t\in X_{\textrm{NE}},\quad \textrm{almost surely,}$$  
where based on the randomization scheme almost sure convergence to different Nash equilibria can be achieved.
\end{myTheo}

Note that the above theorem implies the convergence of behavior probabilities of the game to exact equilibria. It is also worth mentioning that \texttt{Experimental Regret Testing} is a $T$-recall (and hence finite recall) and stationary rule. By contrast, \texttt{ALERT} is neither stationary nor finite-recall. 

% Note, however, that it is not completely uncoupled . As a result, ALERT is not appropriate for the case of unkown games. Using some modifications, however, one can tailor this procedure for such games which is addressed in the next subsection. 

\subsection{Payoff-Based ALERT}
Although \texttt{ALERT} is an uncoupled rule, it fails the requirement of complete uncoupledness as each player relies on the complete knowledge of her own payoff function during play. 
Here we present some modification to \texttt{ALERT} proposed in \cite{regret_testing_germano} to make it completely uncoupled. This modification is based on the idea presented by Foster and Young in \cite{regret_testing}. 

First we highlight that regret calculation is the sole point that each player needs to have the complete knowledge of her payoff. To obtain a completely uncoupled rule, one approach is to estimate the regret. At each time $t$, player $i$ with some small probability chooses an action uniformly at random, and with one minus this probability chooses $s_i^t$ randomly according to $x_i^t$. 

The formal modification needed to be applied to \texttt{ALERT} is as follows. Each player $i$, draws $g_i$ samples for each $h_i=1,\dots,m_i$, with $g_im_i\ll T$.  
Let $U_{i,\tau}\in\{0\}\cup S_i$ be random variables for the current frame. There are exactly $g_i$ values of $\tau$ such that $U_{i,\tau}=h_i$ and all such configurations are equi-probable. For the remaining $\tau$, $U_{i,\tau}=0$. 
At time $\tau$, player $i$ draws an action $s_i^\tau$ as follows. 
If $U_{i,\tau}=h$, $s_i^{\tau}=h$, and if $U_{i,\tau}=0$, $s_i^{\tau}$ is distributed according to $x_i^\tau$. Then the estimated regret will be
$$\hat r_{t,h}^i=\frac{1}{g_i}\sum_{\tau=t+1}^{t+T}\mathbbm 1_{\{U_{i,\tau}=h\}}\pi_i(h,s_{-i}^{\tau})-\frac{1}{T-m_ig_i}\sum_{\tau=t+1}^{t+T}\mathbbm 1_{\{U_{i,\tau}=0\}}\pi_i(s^{\tau}), \quad h=1,\dots,m_i,$$
and other steps are the same. 

As shown in \cite{LPG},  if the regret of player $i$ during the procedure will be at most $\epsilon$ (with  $\epsilon<\rho$), then the probability that the estimated regret $\hat r_{t,h}^i$ is greater than $\rho$ tends to zero as $T$ grows large. Indeed, based on the convergence result for \texttt{Payoff-based ALERT}, one can obtain the following possibility result for the pointwise convergence of behavior probabilities of the game. 

\begin{myTheo}\cite{regret_testing_germano}
For all $\epsilon>0$, there exists a completely uncoupled learning rule such that the mixed action profiles converge almost surely to a profile $x\in X_{\textrm{NE}}^\epsilon$. Moreover, in the case of generic games, the convergence will be to a mixed action profile $x\in X_{\textrm{NE}}$. 
\end{myTheo}

\paragraph{Robustness to Asynchrony.}
In variants of \texttt{Regret Testing}, it is not clear whether the synchronization requirement between players can be relaxed or not. This might become strictly crucial for the case of \texttt{ALERT}, in particular when 
asynchrony can lead to play of different annealed parameters in the game. For example, a portion of players might play with $(T_l,\rho_l,\lambda_l)$ while others might play with $(T_{l'},\rho_{l'},\lambda_{l'})$ for some $l'\neq l$. As a result, a challenging yet interesting question asks whether a.s. (or some weaker notion of) convergence of the play to Nash equilibria is still possible. Our conjecture implies that \texttt{ALERT} is not robust to asynchrony and as such, this learning rule might be merely understood as a possibility result.

\paragraph{Which Mixed Equilibria?} 
For games with several mixed equilibria, and in particular those with a connected Nash component, we face an interesting question: Which mixed equilibria are more likely to be achieved by variants of \texttt{Regret Testing} The procedure itself does not provide any hint as to how this will work. We instead explored it through simple numerical experiments. Application of \texttt{Experimental Regret Testing} to Entry Deterrence game with parameters $(T=10^4,\rho=0.12,\lambda=10^{-3})$ suggested that the procedure is not able to discriminate between different mixed equilibria. Indeed, once the procedure got stuck in an approximate mixed equilibria neighborhood, there is no chance for other mixed equilibria to be reached. 
This is consistent with an statement in \cite{regret_testing_germano} which says based on the realization of initial randomization, any approximate mixed equilibria can be achieved.

%One important question regarding learning rules discussed so far is the type of mixed equilibria that is achieved (or likely to be achieved). For the case of uncoupled, stationary, $R$-recall rules, it is stated in \cite{HM_2006} that amongst multiple mixed equilibria, the more ``mixed'' an equilibrium is, the higher the probability that the learning rules will converge to it. Simple numerical experiments confirm this statement: for example in the case of Entry-Deterrence game,  regret testing most of the times converges to the neighborhood of $(x_{11}=0,x_{21}=0.5)$ which is ``the most mixed'' equilibrium in this game.   
%For non-stationary rules we are unaware of existence of similar results. For instance, in the case of ALERT, it's suggested that depending on the randomization employed in the procedure, any Nash equilibria might be achievable \cite{regret_testing_germano}. 

\section{Pure Equilibria: Learning Rules}
\label{sec:TEL}
In this section, we present three learning rules that can converge to pure equilibria. The first two rules are naive uncoupled rules that obtain pure equilibria in general games. Then, we present \texttt{Trial-and-Error Learning} \cite{LTE} which is a completely uncoupled rule to find pure equilibria in generic games.

\footnotesize
\begin{center}
\begin{tabularx}{.82\textwidth}{X}\
\ \\
\normalsize
\textbf{Algorithm 5:} \\
\textbf{\texttt{\normalsize A Simple Uncoupled Rule for Pure Equilibria}} \cite{LPG}\\ 
\hline
Strategy for player $i$:\\
\ \\
\textbf{Parameters:} The Boolean variable \textsf{convergence}.\\
\ \\
\textbf{Initialization:}\\ 
Choose a pure action for player $i$ uniformly at random. Set \textsf{convergence}=FALSE and $t=1$.\\
\ \\
\textbf{Loop:}\\
1. If \textsf{convergence}=TRUE then repeat the action played in the last odd period. Otherwise do:\\
\ \\
2. If $t$ is odd, choose an action $s^t_i$ randomly. \\
\ \\ 
3. If $t$ is even, let
\begin{eqnarray}
s^{t}_i=\left \{ \begin{array}{ll}
1 & \qquad \textrm{if } \pi_i(s^{t-1})\geq \max_{k\in S_i}\pi(k,s^{t-1}_{-i})\nonumber\\
2 & \qquad \textrm{otherwise}\nonumber\\
\end{array} \right.
\end{eqnarray}
4. If $t$ is even and all players have played action 1, then set \textsf{convergence}=TRUE.\\
\ \\
5. Set $t=t+1$ and repeat the loop.\\
\hline
\end{tabularx}
\end{center}
\normalsize

\footnotesize
\begin{center}
\begin{tabularx}{.82\textwidth}{X}\
\ \\
\normalsize
\textbf{Algorithm 6:} \\
\normalsize
\textbf{\texttt{Another Simple Uncoupled Rule for Pure Equilibria}} \cite{HM_2006}\\ 
\hline
Strategy for player $i$:\\
\ \\
\textbf{Parameters:} - \\
\ \\
\textbf{Initialization:} \\
For $t=1,2$, choose a pure action for player $i$ uniformly at random. Set $t=3$.\\
\ \\
\textbf{Loop:}\\
1. If $s^{t-1}=s^{t-2}$, i.e. all players have played the same action in the last two periods, and $s^{t-1}_i$ was a best reply to $s^{t-1}_{-i}$, then let $s^t_i=s^{t-1}_i$, i.e. repeat the same play. Otherwise choose $s^{t}_i$ uniformly at random. \\
\ \\
2. Set $t=t+1$ and repeat the loop.\\
\hline
\end{tabularx}
\end{center}
\normalsize

If all players play according to these rules, then a pure Nash equilibrium
is eventually achieved, almost surely. This implies that almost every play path consists of a pure Nash equilibria being played from some point on \cite{HM_2006}. Also note that both rules are 2-recall (and hence finite-recall); however, only the second one is stationary. 

\paragraph{Which equilibria?} 
Now that almost sure convergence of play to a pure equilibrium is possible using the above mentioned uncoupled rules, one may ask: in the case of multiplicity of pure equilibria, which one is more likely to be achieved by these procedures? Clearly, using these rules, convergence of play to a pure equilibrium is based on random play. Hence, in case of having multiple pure equilibria, based on the realization of random play any of them might be the convergent action profile in an equi-probable way. Therefore, the two simple procedures presented above cannot distinguish between or be biased towards any of pure equilibria. As a result, they are blind to discriminate between particular kinds of pure equilibria.

\subsection{Trial-and-Error Learning}
\texttt{Trial-and-Error Learning} is a completely uncoupled learning rule that can reach pure  equilibria in generic games  \cite{LTE}. Here we describe \texttt{Trial-and-Error Learning} proposed by Young in \cite{LTE}. A slightly modified variant of this rule is also presented by Pradelski and Young in \cite{LTE_pradelski} that amongst multiple pure equilibria obtains the one that maximizes the welfare of the game.

First we describe \texttt{Trial-and-Error Learning}. 
At a given time moment, the state of player $i$ is the triple $z_i=(\mathcal M_i, \overline{s}_i, \overline{\pi}_i)$, where $\overline{s}_i$ is the current benchmark action of $i$, $\overline{\pi}_i$ is the current benchmark payoff of $i$, and $\mathcal M_i$ is the current mood of $i$, with possible values being \textit{content, discontent, hopeful, and watchful}, respectively symbolized by $c, d, h,$ and $w$ in the sequel. 
Also assume that $\phi(a,b)$ is a function which is monotone increasing in $a$ and monotone decreasing in $b$. Then, \texttt{Trial-and-Error Learning} rule for player $i$ can be described as follows. To proceed, we indeed consider player $i$ in different moods and describe possible state transitions. 
\begin{enumerate}
\item When \textbf{content} $(\mathcal M_i=c)$, the player tries new actions at each time with probability $\epsilon$ (which yields action $s_i$ with payoff $\pi_i(s)$ for her), and with probability $1-\epsilon$ she continues playing with $\overline{s}_i$. Then she changes her state as follows 
\begin{itemize}
\item $z_i=(c,\overline{s}_i, \overline{\pi}_i)\quad$ if $s_i\neq \overline{s}_i$, $\pi_i(s)\leq \overline{\pi}_i$, 
\item $z_i=(c,{s}_i, \pi_i(s))\quad$ if $s_i\neq \overline{s}_i$, $\pi_i(s)> \overline{\pi}_i$,  
\item $z_i=(w,\overline{s}_i, \overline{\pi}_i)\quad$ if $s_i=\overline{s}_i$, $\pi_i(s)< \overline{\pi}_i$, 
\item $z_i=(c,\overline{s}_i, \overline{\pi}_i)\quad$ if $s_i=\overline{s}_i$, $\pi_i(s)= \overline{\pi}_i$, 
\item $z_i=(h,\overline{s}_i, \overline{\pi}_i)\quad$ if $s_i=\overline{s}_i$, $\pi_i(s)> \overline{\pi}_i$. 
\end{itemize} 
\item When \textbf{watchful} $(\mathcal M_i=w)$, she plays her benchmark action $\overline{s}_i$. At each time, she changes her state as follows  
\begin{itemize}
\item $z_i=(d,\overline{s}_i, \overline{\pi}_i)\quad$ if $\pi_i(s)<\overline{\pi}_i$,  
\item $z_i=(c,\overline{s}_i, \overline{\pi}_i)\quad$ if $\pi_i(s)=\overline{\pi}_i$, 
\item $z_i=(h,\overline{s}_i, \overline{\pi}_i)\quad$ if $\pi_i(s)>\overline{\pi}_i$. 
\end{itemize} 
\item When \textbf{hopeful} $(\mathcal M_i=h)$, she plays her benchmark action $\overline{s}_i$. At each time, she changes her state as follows  
\begin{itemize}
\item $z_i=(w,\overline{s}_i, \overline{\pi}_i)\quad $ if $\pi_i(s)<\overline{\pi}_i$,  
\item $z_i=(c,\overline{s}_i, \overline{\pi}_i)\quad$ if $\pi_i(s)=\overline{\pi}_i$, 
\item $z_i=(c,\overline{s}_i, \pi_i(s))\quad$ if $\pi_i(s)>\overline{\pi}_i$. 
\end{itemize} 
\item When \textbf{discontent} $(\mathcal M_i=d)$, she selects an action uniformly at random and changes her state as follows  
\begin{itemize}
\item $z_i=(c,s_i, \pi_i(s))\quad$ with probability $\phi(\pi_i(s),\overline{\pi}_i)$,  
\item $z_i=(d,\overline{s}_i, \overline{\pi}_i)\quad$  with probability $1-\phi(\pi_i(s),\overline{\pi}_i)$.
\end{itemize} 
\end{enumerate}   

Now, we investigate the convergence of \texttt{Trial-and-Error Learning}. First we define the notion of interdependence in normal form games. 

\begin{myDef}\cite{LTE}
A game $G$ is said to be interdependent if any subset $K\subset N$ of players can influence the payoff of at least one player not in $K$, or equivalently 
$$\exists i\notin K,\quad \exists s^\prime_{K}\neq s_{K}\quad\textrm{such that}\quad \pi_i(s^\prime_K,s_{-K})\neq \pi_i(s_K,s_{-K}),$$
where $s_K$ is the restriction of $s$ to the set $K$. 
\end{myDef}

Interestingly, for generic games interdependence holds. Note, however, that interdependence can still hold in games with some payoff ties (and hence non-generic) and therefore, interdependence is a less demanding requirement.
 
\begin{myTheo} \cite{LTE}
Let $G$ be an interdependent $n$-player  game that has at least one pure Nash equilibrium. If players use 
\emph{\texttt{Trial-and-Error Learning}} with experimentation probability $\epsilon$ and function $\phi$, then for all sufficiently small $\epsilon$, \emph{\texttt{Trial-and-Error Learning}} rule converges to pure equilibria with frequency $1-\epsilon$, i.e. pure Nash equilibria will be played at least $1-\epsilon$ of the time. For the case of $n=2$ the interdependence condition can be removed. 
\end{myTheo}

\paragraph{Which equilibria?}
Having presented the convergence result for {\texttt{Trial-and-Error Learning}} rule, we then investigate the answer to this question for multiplicity of pure equilibria. To this effect, we did some numerical experiments for several classical two-player games with multiple equilibria (e.g. Battle-of-the-sexes, Entry Deterrence, Coordination). As all cases yielded the same conclusion, here we only report the result for Entry Deterrence game. Consider Entry Deterrence game with payoff bi-matrix $\bigl(\begin{smallmatrix}
2,2&0,0\\ 1,4&1,4
\end{smallmatrix} \bigr)$,
which has two pure equilibria. Let $P_1$ and $P_2$ respectively denote the pure equilibria with payoff pairs $2,2$ and $1,4$.  

We carried out experiments with $\epsilon=0.01$ and three different choices of function $\phi$, all admitting the format $\phi(a,b)=[p a-q b+c]_x^y$ with $p,q>0$ and $[z]_x^y$ being the projection operator onto $[x,y]$. Table 1 lists frequencies of choosing $P_1$ and $P_2$ averaged over 200 experiments, each lasting $5\times 10^4$ periods. This result shows that using $\phi_1$, both equilibria are almost equi-probable. However, either $P_1$ or $P_2$ is more likely to be achieved if $\phi_2$ or $\phi_3$ has been used. Yet we can construct another function $\phi$ which results in play of $P_1$ with an average frequency greater than those reported in Table 1  
(the statement holds for $P_2$ as well).

\begin{center}
\begin{table}
\footnotesize
\begin{tabular}{ r|c|c|c| }
\multicolumn{1}{r}{}
 &  \multicolumn{1}{c}{$\phi_1(a,b)=$}
& \multicolumn{1}{c}{$\phi_2(a,b)=$} & \multicolumn{1}{c}{$\phi_3(a,b)=$}\\
\multicolumn{1}{r}{}
&  \multicolumn{1}{c}{$[0.001a-0.05b+0.95]_{0.01}^{0.99}$}
& \multicolumn{1}{c}{$[0.6a-0.1b+0.05]_{0.01}^{0.99}$} & \multicolumn{1}{c}{$[0.2a-0.4b+0.15]_{0.01}^{0.99}$}\\
\cline{2-4}
freq. of $P_1$ & 0.471$(1-\epsilon)$  &  0.718$(1-\epsilon)$ & 0.125$(1-\epsilon)$\\
\cline{2-4}
freq. of $P_2$ & 0.529$(1-\epsilon)$  &  0.282$(1-\epsilon)$ & 0.875$(1-\epsilon)$\\
\cline{2-4}
\end{tabular}
\caption{\footnotesize Experimental results for {\texttt{Trial-and-Error Learning}} with $\epsilon=0.01$ in Entry Deterrence game}
\end{table}
\normalsize
\end{center}

It is worthwhile to mention that the work \cite{LTE} only requires $\phi(a,b)$ to be increasing in $a$ and decreasing in $b$. While it provides no other clues for choosing $\phi$, it stresses that convergence is guaranteed under any choice of $\phi$ satisfying these requirements. 
Finally, as our extensive experiments verify, the choice of $\phi$ influences the equilibrium that is likely to be achieved had the game possessed multiple pure equilibria. Therefore, unlike the two uncoupled procedures in the previous subsection, \texttt{Trial-and-Error Learning} is not blind; rather it can be adjusted, perhaps in quite a complicated and tricky way, to converge to a certain equilibrium most of time. 
It, however, remains an interesting open question which format should $\phi$ admit such that convergence to, e.g. a proper equilibrium, to be more likely. 

It should be mentioned, however, that using some modifications, one can obtain a variant of \texttt{Trial-and-Error Learning} such that, in games with multiple pure equilibria, it converges with frequency $1-\epsilon$ to a pure equilibria with the highest social welfare\footnote{The social welfare of action profile $s$ is $\sum_{i\in N} \pi_i(s)$.}. This is indeed the completely uncoupled procedure proposed by Pradelski and Young \cite{LTE_pradelski}, which converges with frequency $1-\epsilon$ to 
$\arg\max_{s\textrm{ is PNE}}\sum_{i\in N}\pi_i(s)$. 

\section{Fictitious Play} 
\label{sec:fic_play}
Perhaps fictitious play is amongst the first uncoupled rules invented for repeated play of games. In order to keep the scope of this report unified, we will not pay too much attention to this rule and its variants. However, as an uncoupled rule, here we will briefly investigate its appeal and its drawback for the sake of completeness.  

In fictitious play, at time $t$, each player $i$ chooses an action that is a best response to empirical distribution up to time $t-1$: 
$$s_i^t=\arg\max_{h\in S_i}\frac{1}{t-1}\sum_{\tau=1}^{t-1}\pi_i(h,s_{-i}^\tau).$$

Fictitious play is not guaranteed to converge to Nash equilibria for general games. Indeed for the general case, even convergence to the set of correlated equilibria may not be achieved using this rule (see e.g. \cite{LPG}). However, for the case of two-player zero-sum games, it has been shown in \cite{Robinson}, that using fictitious play, the joint empirical distributions of play converges to the set of Nash equilibria. 

Within decades, several variants of fictitious play such as \emph{smoothed fictitious play}, \emph{vanishingly smoothed fictitious play} have been proposed (see e.g. \cite{fudenberg_levine,LPG}). In particular, in \cite{fictitious_play_hofbauer}, Hofbauer and Sandholm proposed stochastic fictitious play for which the convergence to Nash equilibria can be guaranteed.

\section{Demarcating Possible and Impossible}
\label{sec:poss_imposs}

As discussed earlier, a fundamental step when dealing with simple uncoupled and completely uncoupled learning rules is to distinguish between what is possible and what is not. 
In the course of the last decade, several works paid attention to explore the border between possibilities and impossibilities for coupled and completely uncoupled learning rules (see e.g. \cite{babi_2012, HM_2013, HM_impossibility, LTE, regret_testing_germano}). Here, we briefly reflect the main results. 

\subsection{Uncoupled Rules}
\paragraph{Pure Nash Equilibria}
\begin{itemize}
\item There exist no uncoupled, 1-recall, stationary learning rules that lead to a.s. convergence of play to pure Nash equilibria for all games (e.g. \cite{HM_2006}, Theorem 1). 

\item There exist uncoupled, 1-recall, stationary rules that can lead to a.s. convergence of play to pure Nash equilibria in all generic two-player games (\cite{HM_2006}, Proposition 2). 

\item There exist uncoupled, 2-recall, stationary rules that can lead to a.s. convergence of play to pure Nash equilibria in all games (\cite{HM_2006}, Theorem 3). One instance of such uncoupled learning rules has been outlined as Algorithm 6.  
\end{itemize}

\paragraph{Mixed Nash $\epsilon$-Equilibria} 
\begin{itemize}
\item For every $M$ and $\epsilon>0$, there exists an integer $R$ and an uncoupled, $R$-recall, stationary rule that can lead to a.s. convergence of empirical distributions of play to Nash $\epsilon$-equilibria in all games. For almost every history of play, there exists a Nash $\epsilon$-equilibrium $x$ such that for every combination of $s\in S$, 
$$\lim_{t\rightarrow\infty}\Phi_t[s]=\prod_{i\in N} x_i(s_i),\quad\hbox{almost surely.}$$
Moreover, there exists an almost surely finite stopping time $T$ after which we have (\cite{HM_2006}, Theorem 5):
$$\lim_{t\rightarrow\infty}\Pr(s^t=s|H_T)=\prod_{i\in N} x_i(s_i),\quad\hbox{almost surely.}$$
Note, however, that it does not imply the convergence of behavior probabilities of the play. 

\item There exists $\epsilon_0$ (dependent on $n$ and $m_i, i\in N$) such that for every $\epsilon<\epsilon_0$, there are no uncoupled, finite recall, stationary rules that lead to a.s. convergence of behavior probabilities to Nash $\epsilon$-equilibria in all games (\cite{HM_2006}, Theorem 6). 

\item For every $\epsilon>0$, there exist integer $R$ and an uncoupled, $R$-memory, stationary rule that leads to a.s. convergence of behavior probabilities of play to Nash $\epsilon$-equilibria in all games.  (\cite{HM_2006}, Theorem 7)  
\end{itemize}

\subsection{Completely Uncoupled Rules}

\paragraph{Pure Nash Equilibria}
\begin{itemize}

\item There are no completely uncoupled rules that lead to a.s. convergence of play to pure Nash equilibria, in all games. This statement holds true even if we restrict the attention to generic games  (\cite{babi_2012}, Corollary 1).

\item There are no completely uncoupled rules that lead to play of pure Nash equilibria in all games, even for convergence with frequency $1-\epsilon$ (\cite{babi_2012}, Theorem 3). For the latter, however, for all two-player games, there exists such a rule: \texttt{Trial-and-Error Learning} \cite{LTE}. 

\item There is a completely uncoupled rule that leads to play of pure Nash equilibria in all interdependent (and hence generic) games, with frequency $1-\epsilon$. The only known rule satisfying this statement \texttt{Trial-and-Error Learning} \cite{LTE}.
 
\item Under some (uncoupled) additional information, that include either the knowledge of the number of players $n$ or the index of each player $i$, there exists a completely uncoupled rule that leads to a.s. convergence of play to pure Nash equilibria in all generic games (\cite{babi_2012}, Theorem 4).

\item Even under (uncoupled) additional information described above, there are no completely uncoupled rules that lead to play of pure Nash equilibria with frequency $1-\epsilon$ in all games (\cite{babi_2012}, Corollary 3). 

\end{itemize}

\paragraph{Mixed Nash $\epsilon$-Equilibria} 
\begin{itemize}
\item There exist completely uncoupled learning rules that lead to a.s. convergence of behavior probabilities of play to an $\epsilon$-equilibrium in every generic game (\cite{babi_2012}, Theorem 5). One such rule is \texttt{Payoff-based ALERT} \cite{regret_testing_germano}. 

\item There exist completely uncoupled rules that lead to convergence of behavior probabilities of play to an $\epsilon$-equilibrium in all games with frequency $1-\epsilon$ (\cite{babi_2012}, Theorem 6)\footnote{It is unclear whether that \texttt{Payoff-based ALERT} satisfies this requirement \cite{regret_testing_germano}.}. 

\item There are no completely uncoupled rules realizable with finite-memory that lead to a.s. convergence of behavior probabilities of play to $\epsilon$-equilibria in all games (\cite{babi_2012}, Theorem 7).  
 
\end{itemize}

We note that pure Nash equilibria are exact points. By contrast a mixed Nash $\epsilon$-equilibrium is not an exact point; indeed it constitutes a set. Therefore, if occasionally we reach a pure Nash equilibrium,  using completely uncoupled rules it is impossible to stay there thereafter. On the other hand, it is possible to get stuck in mixed Nash $\epsilon$-equilibria while players are still searching.

Tables 2 and 3 summarize the results presented in this section. The identifier `N/A' denotes open issues. 

\begin{center}
\begin{table}
\footnotesize
\begin{tabular}{ r|c|c|c|c| }
\multicolumn{1}{r}{}
 &  \multicolumn{1}{c}{uncoupled}
& \multicolumn{1}{c}{uncoupled}
 & \multicolumn{1}{c}{completely uncoupled} 
& \multicolumn{1}{c}{completely uncoupled}\\
\multicolumn{1}{r}{}
& \multicolumn{1}{c}{(1-recall,} & \multicolumn{1}{c}{($R^{\geq 2}$-recall,} & \multicolumn{1}{c}{} & \multicolumn{1}{c}{(with additional}\\
\multicolumn{1}{r}{}
& \multicolumn{1}{c}{stationary)} & \multicolumn{1}{c}{stationary)} & \multicolumn{1}{c}{} & \multicolumn{1}{c}{information)}\\
\cline{2-5}
\multirow{2}{*}{a.s. conv.} &  all: \ding{55} &  all: \ding{51} & all: \ding{55} & all: \ding{55}\\
\cline{2-5}
& generic ($n=2$): \ding{51} & - & generic: \ding{55} & generic: \ding{51} \\
\cline{2-5}
\multirow{2}{*}{freq. $1-\epsilon$}   &  all: N/A &  all: \ding{51}  & all: \ding{55} & all: \ding{55} \\
\cline{2-5}
& generic: N/A & - & generic: \ding{51} & generic: \ding{51} \\
\cline{2-5}
\end{tabular}
\caption{\footnotesize Summary of possibilities and impossibilities for pure Nash equilibria}
\normalsize
\end{table}
\end{center}

\begin{center}
\begin{table}
\footnotesize
\begin{tabular}{ r|c|c|c| }
\multicolumn{1}{r}{}
 &  \multicolumn{1}{c}{uncoupled}
& \multicolumn{1}{c}{uncoupled}
& \multicolumn{1}{c}{completely}\\
\multicolumn{1}{r}{}
& \multicolumn{1}{c}{(finite-recall)} & \multicolumn{1}{c}{(finite-memory)} 
& \multicolumn{1}{c}{uncoupled}\\
%\multicolumn{1}{r}{}
%& \multicolumn{1}{c}{-} & \multicolumn{1}{c}{-} & \multicolumn{1}{c}{} & \multicolumn{1}{c}{information)}\\
\cline{2-4}
\multirow{2}{*}{behavior (a.s. conv.)} &  all: \ding{55} &  all: \ding{51} & all: \ding{55}\\
\cline{2-4}
& generic: N/A & -  & generic: \ding{51} \\
\cline{2-4}
\multirow{2}{*}{distribution (a.s. conv.)}   &  all: \ding{51} &  all: \ding{51} & all: \ding{55} \\
\cline{2-4}
& - & - & generic: \ding{51} \\
\cline{2-4}
\multirow{2}{*}{distribution (freq. $1-\epsilon$)}   &  all: \ding{51} &  all: \ding{51}  & all: \ding{51} \\
\cline{2-4}
& - & - & -\\
\cline{2-4}
\end{tabular}
\caption{\footnotesize Summary of possibilities and impossibilities for approximate mixed equilibria}
\normalsize
\end{table}
\end{center}

\subsection{Time Efficiency}
Here we briefly address complexity of reaching equilibria using uncoupled rules. In general we are interested in \emph{time-efficient} learning rules, where time efficiency is defined as follows: A learning rule is time-efficient if the time it takes to converge to equilibria is polynomial in the
number of players.

If we were to find correlated $\epsilon$-equilibria of the game, the communication complexity would be \cite{LPG}: $$\max_{i\in N} \frac{16m_in}{\epsilon^2}\log\frac{m_in}{\delta}\quad \hbox{with probability at least $1-\delta$},$$
which is polynomial in game parameters (i.e. $n$ and $m_i,i\in N$) and is proportional to $\frac{1}{\epsilon^2}$. This implies that approximate correlated equilibria for every finite game can be found using time-efficient uncoupled rules. 
This is in sharp contrast to the fact that as of today, no polynomial-time algorithm is known to find approximate Nash equilibria in every game \cite{papadimitriou, comcomplexity}.   
In particular, for communication complexity of uncoupled rules to find pure equilibria, we have the following impossibility result.

\begin{myTheo}\cite{comcomplexity}
There are no time-efficient uncoupled learning rules that reach a pure Nash equilibrium in all games where such equilibria exist.
\end{myTheo}

Comparing the communication complexity of reaching approximate correlated equilibria and the assertion of theorem above motivates us to further explore such a dramatic difference in communication complexity of reaching correlated and Nash equilibria.
As postulated by Hart \cite{HM_2013}, this is in line with the fact that correlated equilibria are solutions of linear inequalities whereas Nash equilibria are fixed points of nonlinear mappings. One conclusion is that correlated equilibria is a ``dynamically easy'' concept while Nash equilibria is a ``dynamically hard'' one. This conclusion along with impossibility results outlined above inherently implies \emph{the law of conservation of coordination,} which was coined perhaps by Sergiu Hart: In a general setup, the need for coordination cannot be removed; it should be embedded in either the solution concept (as in the case for correlated equilibria) or the underlying learning rule.

\section{Discussion}
\label{sec:discuss}
This section is devoted to discuss some pitfalls of learning rules discussed so far. 

\paragraph{Which Equilibria?} 
One important question regarding learning rules discussed so far is the type of equilibria that is achieved (or is more likely to be achieved) had the underlying game possessed multiple equilibria. 
We addressed this question in earlier sections. To summarize, 
\begin{itemize}
\item For learning pure equilibria using \texttt{Trial-and-Error Learning}, the choice of function $\phi$ proves capable of influencing the learning rule to decide in favor of one equilibria most of the time. However, it is unclear how to choose $\phi$ such that a perfect or proper equilibrium to be selected most of the time (and in all games). 

\item For the case of mixed equilibria, regret based learning procedures discussed so far are blind in distinguishing between different (approximate) mixed equilibria. All approximate mixed equilibria are attainable in an equi-probable way. 
\end{itemize}

%For the case of uncoupled, stationary, $R$-recall rules, it is stated in \cite{HM_2006} that amongst multiple mixed equilibria, the more ``mixed'' an equilibrium is, the higher the probability that the learning rules will converge to it. Simple numerical experiments confirm this statement: for example in the case of Entry-Deterrence game,  regret testing most of the times converges to the neighborhood of $(x_{11}=0,x_{21}=0.5)$ which is ``the most mixed'' equilibrium in this game.   
%For non-stationary rules we are unaware of existence of similar results. For instance, in the case of ALERT, it's suggested that depending on the randomization employed in the procedure, any Nash equilibria might be achievable \cite{regret_testing_germano}. 

\paragraph{Pareto Efficiency.} 
Learning rules discussed so far do not imply whether an efficient profile\footnote{In terms of the social welfare function of the game} will be reached. Achieving efficient action profiles is an issue of great concern when considering system level objectives and has great implications in engineering disciplines. 
To our best knowledge, the only completely uncoupled learning rule that guarantees the convergence to an efficient pure equilibria is a variant of \texttt{Trial-and-Error Learning} presented by Pradelski and Young \cite{LTE_pradelski}. However, we are unaware of the existence of uncoupled learning rules that converge to efficient mixed action profiles.    

\paragraph{Susceptibility to Delayed and Inexact Observations, and Asynchrony.} Both uncoupled and completely uncoupled learning rules are appropriate models for decentralized decision making. An inherent feature of such systems is that each player is informed of her payoff after some delay usually modeled as a random process. Another inherent feature is the possibility of  inexact information, which means that players might receive noisy observation sequence. 

One interesting question that may arise is: For the case of possible results, is it still possible to reach equilibria using delayed payoffs? Is it possible to allow players to take their actions based on inexact observations? 
To the best of our knowledge, the majority of previous studies dealt with players being informed of their payoffs in a delay-free and noise-free manner. The only closely related  work is perhaps the working paper  by Lagziel and Lehrer \cite{delayed_info}, in which they address the possibility of reaching (correlated) equilibria under some restrictions for delay. Another distantly related work might be the work by Sandholm \cite{inexact}.    

Another issue which is relevant for the case of learning procedure that has a frame structure (in particular \texttt{ALERT}) is susceptibility to asynchrony. In particular, for \texttt{ALERT} asynchrony can lead to play of different annealed parameters in the game. For example, a portion of players might play with $(T_l,\rho_l,\lambda_l)$ while others that joined later to the process might play with $(T_{l'},\rho_{l'},\lambda_{l'})$ for some $l'\neq l$. As a result, a challenging yet interesting question asks whether a.s. (or some weaker notion of) convergence of the play to Nash equilibria is still possible. Our conjecture implies that \texttt{ALERT} is not robust to asynchrony and as such, this learning rule might be merely understood as a possibility result.

%\paragraph{Remark 3.} In the normal form games considered thus far, we assumed that the absolute value of payoffs do not exceed some constant $M$. 
%Although rules like trial and error learning and variants of regret testing do not require $M$ to be known by the players, the assertions of some possibility results presented so far inherently rely on players who are aware of this upper bound. In light of techniques like annealing it seems, however, plausible to remove the dependence of $M$. In this regard, some similar techniques can be also borrowed from bandit theory. 

\paragraph{Exploiting Payoff Structure.} 
All learning rules discussed above are quite close to exhaustive search. As a result, their worst case convergence time (for the general case) grows exponentially as a function of parameters involved in the game. With the exception of potential games, it is not clear whether specific mathematical structures in the payoff functions involved (like supermodularity) will prove helpful in devising more efficient learning rules. 

%This issue is relevant especially in the case of continuous games, e.g. those with compact set of actions. The work \cite{CorrEq_Compact} deals with learning correlated equilibria using uncoupled rules in games with compact set of actions. For the case of Nash equilibria, however, no such result is available and the existing works exploited the mathematical structure in payoffs only to facilitate the computation of Nash equilibria and just for particular classes of games. 

%This is indeed in line with the impossibility results in \cite{comcomplexity} stated in Section \ref{sec:poss_imposs}. Is there some specific class of game for which time efficient (i.e. polynomial time) rules exist? Is there some $\epsilon_0$ such that Nash $\epsilon$-equilibria with $\epsilon> \epsilon_0$ can be achieved in a time efficient way? 
%Indeed the work by Daskalakis, Mehta, and Papadimitriou \cite{papadimitriou} (and also references therein) provides partially positive answer to this question for the case of two-player games. However, since the scope of \cite{papadimitriou} is different  from this work, we didn't address it here. 

\paragraph{Being Unreasonable.}
It is mentioned earlier that cognitive  optimization degree requirement of learning rules of interest is greater than that of evolutionary dynamics and less than that of fully rational learning. One drawback associated to \emph{our} learning rules is that they are not reasonable, rationalizable, etc. They just try to implement a search procedure that is a bit smarter than exhaustive search. More importantly, they do not try to exploit the correlation between players through the payoff function. As such they fail to reflect realistic decision process of players with bounded rationality. 

It seems promising to explore rules that lie between \emph{our} learning rules and fully rational learning in the spectrum of cognitive optimization of learning procedures. It seems that these smarter rules to be capable of improving upon serious drawbacks of our rules, and consequently to be able to capture decision making of boundedly rational players more precisely. 

\paragraph{Connection to Information Theory} 
We can think of drawbacks and impossibility results of both uncoupled and completely uncoupled learning rules as a result of bounded rationality. In particular, recall in Section \ref{sec:poss_imposs} where we restricted the uncoupled rules to be finite-memory or even finite-recall. These are inherent restrictions of players of players with bounded rationality. Although some information-theoretic results for non-cooperative game theory with players with bounded rationality have been established in \cite{infotheory}, as of today 
no result targeting information-theoretic impossibility of uncoupled learning rules has been observed. One possible road is to investigate impossibilities in light of information theory though it sounds like a hope.

%\paragraph{Engineering Implications}
%Here we stress some possible implications of learning rules discussed so far for engineering disciplines. Nowadays, learning equilibria in decentralized systems is an issue of great concerns because of both emergence of many decentralized application and widespread and ever increasing use of game theory to model these problems. A plethora of reasons, which we cannot afford to cover here, restricts decentralized decision making to fall in the domain of uncoupled and completely uncoupled learning rules. 
%Indeed, the implications of such dynamics is relevant in the case of investigation of such dynamics for particular classes of games, such as potential games, congestion games, supermodular games, to name a few. While the cutting power of dynamics presented so far is limited (in terms of attainable convergence, convergence speed), it is possible to devise dynamics of the same type to exploit the structure in the payoff and game. 
%
%While it is possible to directly address , the line of study proposed here clarifies the border of possibilities and impossibilities, particularly when extended to special classes of games. This will serve as a certificate to prove the optimality or to denote some gap in the suboptimality of realistic dynamics to achieve equilibria in engineering applications. 

\subsection{Lessons Learned}
So far, we have considered several uncoupled and completely uncoupled learning rules that can reach different types of equilibria in \emph{almost} all finite games. The essential concern of this paper, however, was not to provide a list of such rules. Rather, our first concern was to address the possibility of reaching (Nash or correlated) equilibria using (possibly simple) learning rules that rely on players with limited observability (i.e. uncoupledness and completed uncoupledness). Second, it aimed to present the solution concepts upon which such learning rules were built.  

For correlated equilibria, we have seen the existence of time-efficient uncoupled and completely uncoupled learning rules that work in every finite game. The discussion in the previous sections also showed that such rules have an appealing communication complexity as a function of number of players and their pure actions. As for their underlying solution concept, the notion of \emph{internal regret} proved valuable. 

By contrast, it was understood that there exist no uncoupled learning rules that guarantee almost sure convergence of play of Nash equilibria in all finite games. Some refinements of the main question then emerged: Are there uncoupled learning rules that guarantee almost sure convergence to Nash equilibria in all generic games? Are there uncoupled learning rules to obtain convergence  to Nash equilibria with frequency $1-\epsilon$ in all games? 

Although the answers to many questions like these are negative (e.g., results in Section \ref{sec:poss_imposs}), some works answered to some of them in the affirmative by providing appropriate rules (e.g. \cite{regret_testing_germano, LTE}). 
Note, however, that for the case of positive answers, in a plethora of situations such rules do not exhibit elegant convergence behavior and as of today they might be considered solely as possibility results.

\section*{\centering{Acknowledgment}}
The author would like to thank J\"{o}rgen Weibull for his insightful comments on an earlier version of this paper.

%%%%%%%%%%%%%%%%%%%%%%%%%%%%%%%%%%%%%%%%%%%%%%%%%%%%%%%%%%%%%%%%%%%%%%%%%%%%%%%%%%%%%%%%%%%%%%%%%%%%%%%%%
%%%%%%%%%%%%%%%%%%%%%%%%%%%%%%%%%%%%%%%%%%%%%%%%%%%%%%%%%%%%%%%%%%%%%%%%%%%%%%%%%%%%%%%%%%%%%%%%%%%%%%%%%
%%%%%%%%%%%%%%%%%%%%%%%%%%%%%%%%%%%%%%%%%%%%%%%%%%%%%%%%%%%%%%%%%%%%%%%%%%%%%%%%%%%%%%%%%%%%%%%%%%%%%%%%%


\begin{thebibliography}{99}

\bibitem{nonstochastic_bandit}
Auer, P., Cesa-Bianchi, N., Freund, Y., Schapire, R.E., 2002. The nonstochastic multiarmed bandit problem. SIAM J. Comput. 32, 48–-77.

\bibitem{babi_2010}
Babichenko, Y., 2010. Uncoupled automata and pure Nash equilibria. Int. J. Game Theory. 39, 483-–502.

\bibitem{babi_2012}
Babichenko, Y., 2012. Completely uncoupled dynamics and Nash equilibria. Games Econ. Behav. 76, 1--12.


\bibitem{blume}
Blume, L., 1993. The statistical mechanics of strategic interaction. Games Econ. Behav. 5, 387–424.

\bibitem{fictitious_play}
Brown, G.W., 1951. Iterative solution of games by fictitious play. Activity Analysis of Production and Allocation. 374--376. 

\bibitem{LPG}
Cesa-Bianchi, N., Lugosi, G., 2006. Prediction, Learning, and Games. Cambridge Univ. Press.

\bibitem{papadimitriou}
Daskalakis C., Mehta, A., Papadimitriou, C., A note on approximate Nash equilibria. Theo. Comp. Sci. 410, 1581--1588. 


%\bibitem{hypo_testing}
%Foster, D.P., Young, H.P., 2003. Learning, hypothesis testing, and Nash equilibrium. Games Econ. Behav. 45, 73--96.



\bibitem{regret_testing}
Foster, D.P., Young, H.P., 2006. Regret-testing: Learning to play Nash equilibrium without knowing you have an opponent. Theoretical Econ. 1, 341--367.

\bibitem{nash_seeking_2}
Frihauf, P., Krstic, M., Basar, T., 2012. Nash Equilibrium Seeking in Noncooperative Games.
IEEE Trans. Auto. Control 57(5), 1192--1207.

\bibitem{fudenberg_levine}
Fudenberg, D., Levine, D., 1998. The theory of learning in games. Cambridge: MIT Press.

\bibitem{regret_testing_germano}
Germano, F., Lugosi, G., 2007. Global Nash convergence of Foster and Young's regret-testing. Games Econ. Behav. 60, 135-–154.

%Hart, S., 2011. Nash equilibrium and dynamics. Games Econ. Behav. 71, 6–8.


\bibitem{HM_simple_adaptive}
Hart, S., Mas-Colell, A., 2000. A simple adaptive procedure leading to correlated equilibrium. Econometrica 68, 1127–-1150.

\bibitem{HM_reinforcement_learning}
Hart, S., Mas-Colell, A., 2001. A Reinforcement Procedure Leading to Correlated Equilibrium. Springer Economic Essays. 181--200. 

\bibitem{HM_impossibility}
Hart, S., Mas-Colell, A., 2003. Uncoupled dynamics do not lead to Nash equilibrium. Amer. Econ. Rev. 93, 1830–-1836.

\bibitem{HM_2006}
Hart, S., Mas-Colell, A., 2006. Stochastic uncoupled dynamics and Nash equilibrium. Games Econ. Behav. 57, 286–-303.

\bibitem{comcomplexity}
Hart, S., Mansour, A., 2010. How long to equilibrium? The communication complexity of uncoupled equilibrium procedures. Games Econ. Behav. 69, 107–-126.

\bibitem{HM_2013}
Hart, S., Mas-Colell, A., 2013 (forthcoming). Simple Adaptive Strategies: From Regret-Matching to Uncoupled Dynamics. World Scientific Publishing.

%Neyman, A., 1998. Finite repeated games with finite automata. Math. Oper. Res. 23, 513–552.
%Young, H.P., 2007. The possible and the impossible in multi-agent learning. Artificial Intell. 171, 429–433.

\bibitem{fictitious_play_hofbauer}
Hofbauer, J., Sandholm, W.H., 2002. On the Global Convergence of Stochastic Fictitious Play. Econometrica 70(6), 2265--2294.

\bibitem{delayed_info}
Lagziel, D., Lehrer, E., 2012. Regret-free strategy with delayed information. Discussion paper, School of Mathematical Sciences, Tel Aviv University.

\bibitem{krishna}
Maskery, M., Krishnamurthy, V., Zhao, Q., 2009. Decentralized dynamic spectrum access for cognitive radios: cooperative design of a non-cooperative game. IEEE Trans. Comm. 5(2), 459--469.  

\bibitem{loglinear}
Marden, J.R., Shamma, J.S., 2012. Revisiting log-linear learning: asynchrony, completeness and payoff-based implementation. Games Econ. Behav. 75, 788–-808.

\bibitem{LTE_pradelski}
Pradelski, B.S.R., Young, H.P., 2012. Learning efficient Nash equilibria in distributed systems. Games Econ. Behav. 75, 882–-897.

\bibitem{Robinson}
Robinson, J., 1951. An iterative method of solving a game. The Annals of Mathematics, 54(2), 296--301.

\bibitem{inexact}
Sandholm, W.S., 2003. Evolution and Equilibrium under Inexact Information. Games Econ. Behav. 44, 343--378.

\bibitem{nash_seeking}
Stankovic', M.S., Johansson, K.H., Stipanovic', D.M., 2012. Distributed Seeking of Nash Equilibria with Applications to Mobile Sensor Networks. IEEE Trans. Automatic Control, 57(4), 904--919.

%\bibitem{CorrEq_Compact}
%Stoltz, G., Lugosi, G., 2007. Learning correlated equilibria in games with compact set of strategies. Games Econ. Behav. 59, 187--208. 


\bibitem{infotheory}
Wolpert, D.H., 2004. Information theory-- the bridge connecting bounded rational game theory and statistical physics. Complex Engineering Systems, D. Braha and Y. Bar-Yam (Ed.'s).


\bibitem{LTE}
Young, H.P., 2009. Learning by trial-and-error. Games Econ. Behav. 65, 626-–643.

\bibitem{Strategic_learning}
Young, H.P., 2004. Strategic Learning and Its Limits. Oxford Univ. Press. 


\end{thebibliography}
\end{document}